$J_1$–$J_2$ Square-Lattice Heisenberg Antiferromagnets with $4d^1$ spins: $A$MoOPO$_4$Cl ($A$ = K, Rb)

[1]Hajime Ishikawa, [1]Nanako Nakamura, [1]Makoto Yoshida, [1]Masashi Takigawa, [2]Peter Babkevich, [3]Navid Qureshi, [2]Henrik M. Rønnow, [1]Takeshi Yajima, [1]Zenji Hiroi

[1]Institute for Solid State Physics (ISSP), The University of Tokyo, Kashiwa, Chiba, 277-8581, Japan

[2]Laboratory for Quantum Magnetism, Institute of Physics, École Polytechnique Fédérale de Lausanne (EPFL), CH-1015 Lausanne, Switzerland

[3]Institut Laue-Langevin, BP 156, F-38042, Grenoble Cedex 9, France

Magnetic properties of $A$MoOPO$_4$Cl ($A$ = K, Rb) with Mo$^{5+}$ ions in the $4d^1$ electronic configuration are investigated by magnetization, heat capacity and NMR measurements on single crystals, combined with powder neutron diffraction experiments. The magnetization measurements reveal that they are good model compounds for the spin-1/2 $J_1$–$J_2$ square lattice magnet with the first and second nearest-neighbor interactions. Magnetic transitions are observed at around 6 and 8 K in the K and Rb compounds, respectively. In contrast to the normal Néel-type antiferromagnetic order, the NMR and neutron diffraction experiments find a columnar antiferromagnetic order for each compound, which is stabilized by a dominant antiferromagnetic $J_2$. Both compounds realize the unusual case of two interpenetrating $J_2$ square lattices weakly coupled to each other by $J_1$.

I. INTRODUCTION

Low-dimensional quantum magnets show rich ground states and phase transitions owing to strong quantum fluctuations. The spin-1/2 $J_1$–$J_2$ square lattice magnet, where spins on the square lattice are coupled by the nearest neighbor (NN) interaction $J_1$ and the next nearest neighbor (NNN) interaction $J_2$ in the diagonal direction, is one of the interesting examples. For classical spins, three ground states appear depending on the sign and the relative magnitude of $J_1$ and $J_2$ [1]; the $J_1$–$J_2$ phase diagram is illustrated as a function of the $J_1$ and $J_2$ in Fig. 1. When $J_1$ is dominant ($|J_1| > 2J_2$), Néel antiferromagnetic (NAF) and ferromagnetic (F) states are realized for antiferromagnetic ($J_1 > 0$) and ferromagnetic ($J_1 < 0$) couplings, respectively, as in the case of a conventional spin-1/2 square lattice magnet with $J_2 = 0$. On the other hand, when antiferromagnetic $J_2$ is dominant ($|J_1| < 2J_2$), columnar antiferromagnetic (CAF) state with a stripe spin structure is stabilized. Moreover, strong quantum fluctuations for spin 1/2 add interesting features to the $J_1$–$J_2$ phase diagram. Novel quantum ground states without magnetic long-range order are predicted to occur in the highly frustrated parameter regions with $|J_1| \sim 2J_2$, although the details of these states are still under debate. A spin liquid state [2] or plaquette valence bond phases [3] are predicted for $J_1 > 0$, while a spin nematic state, which is related to the Bose-Einstein condensation of bound magnon pairs, is predicted for $J_1 < 0$ [4]. The $J_1$–$J_2$ model is also important as it is likely relevant to the mechanism of the high-temperature superconductivity of the iron pnictides [5,6].

Typical model compounds for the convensional spin-1/2 square lattice magnet are found in cuprates with $Cu^{2+}$ ($3d^9$) ions. For example, magnetic properties of $La_2CuO_4$ [7,8], which is known as the parent compound of the high-temperature superconductivity, $Sr_2CuO_2Cl_2$ [9,10], $Cu(DCOO)_2·4D_2O$ [11], $Cu(pyz)_2(ClO_4)_2$ [12], and $[Cu(pyz)_2(HF_2)]BF_6$ [13] have been investigated. In these compounds, $J_1$ is largely antiferromagnetic, while $J_2$ is almost negligible. In the case of the oxides, strong hybridizations between Cu $d_{x2-y2}$ and oxygen $p$ orbitals cause dominant $J_1$. As for model compounds for the spin-1/2 $J_1$–$J_2$ square lattice magnet, various vanadates with $V^{4+}$ ($3d^1$) ions have been studied (Fig.1). Although these vanadates have spins in the V $d_{xy}$ orbitals in the compressed $VO_5$ square pyramidal coordination in common, the signs and magnitudes of $J_1$ and $J_2$ differ from one material to another. The first experimental realization of the spin-1/2 $J_1$–$J_2$ square lattice model was found in $Li_2VOMO_4$ ($M$ = Si, Ge), in which the CAF order is observed due to the weakly antiferromagnetic $J_1$ and strongly antiferromagnetic $J_2$ [14-16]. CAF orders are also observed in $AA'VO(PO_4)_2$ ($AA'$ =BaCd, SrZn, BaZn, $Pb_2$, PbZn), which have similar layered structures to $Li_2VOMO_4$, while $J_1$ is weakly ferromagnetic [19-22]. In these compounds, the peculier layered structure may reduce $J_1$ and enhance $J_2$. In other vanadates such as $VOMoO_4$ [17] and $PbVO_3$ [18], dominant antiferromagnetic $J_1$ stabilizes NAF. Unfortunately, no compounds seem to exist in the boundary regimes in the phase diagram. Therefore, in order to explore the quantum phases theoretically predicted in the boundary regimes in real materials, it is necessary to search for new $J_1$–$J_2$ magnets having appropriate $J_1$–$J_2$ ratio.

From the view point of the material exploration, expanding the target from $3d$ toward $4d$/$5d$ electron compounds would be attracting. Although less examples of localized spin systems have been found in the $4d$ compounds, an early study indicated that the fluorides $AMoF_6$ (A = Na, K, Rb, Cs) are good examples for localized $4d$ spin systems [23]. In recent years, more materials which show unique magnetic properties are reported. For example, the perovskite oxide $SrTcO_3$ shows an unusually high Néel temperature [24], which is considered to be caused by the large hybridization between the spatially extended $4d$ orbitals and oxygen $p$ orbitals. Such extended $4d$ orbitals may be utilized to realize a different balance between the NN and NNN magnetic interactions in the square lattice compared with those in the $3d$ systems. Moreover, the possible realizations of magnetic models that are different from the conventional Heisenberg model are discussed in $4d$ electron systems such as the molybdenum double perovskite oxides [25] and $\alpha$-$RuCl_3$ [26]. From these aspects, we expect that there is a chance to find a new $J_1$–$J_2$ square lattice magnet in $4d$ electron systems, which may lead to the finding of novel magnetic phases theoretically predicted or unpredicted in the $J_1$–$J_2$ square lattice Heisenberg model.

The $4d^1$ compounds $AMoOPO_4Cl$ ($A$ = K, Rb) were reported by Borel et al. in 1998 [27]. They crystalize in the tetragonal space group $P4/nmm$ at room temperature with the lattice constants of $a$ = 6.4340(5) Å and $c$ = 7.2715(9) Å for the K compound and $a$ = 6.4551(8) Å and $c$ = 7.4612(8) Å for the Rb compound. The $Mo^{5+}$ ions with the $4d^1$ electronic configuration should carry spin-1/2, when the spin-orbit coupling is negligible. As shown in the crystal structure in Fig. 2 (a), the $MoO_5Cl$ octahedra and $PO_4$ tetrahedra share their corners to form layers which are separated by K or Rb ions. Every layer consists of a pair of Mo sheets at different heights along the $c$ axis. The Mo atoms are connected linearly via one $PO_4$ tetrahedron within one sheet, while, between two sheets, they are connected via two $PO_4$ tetrahedra at nearly orthogonal angles: the former Mo pair corresponds to the next-nearest-neighbors and the latter to the nearest-neighbors. Superexchange interactions are expected between the NN and NNN spins through the two oxygen ions of $PO_4$ tetrahedra. Therefore, it can be viewed that two square lattices made of $J_2$ are stacked in a staggered way and are coupled by $J_1$ to form $J_1$–$J_2$ square lattice in the structure. The atomic distances between the NN, NNN, and inter-plane Mo atoms are 4.75 (4.75) Å, 6.43 (6.46) Å, and 7.27 (7.46) Å in the K- (Rb-) compounds. In addition to the long atomic distances, the absence of the Mo-O-O-Mo path through the $PO_4$ tetrahedra for superexchange interactions along the inter-plane direction would reduce the inter-plane couplings and make the compounds quasi-two-dimensional. This layered structure resembles the vanadates $Li_2VOMO_4$ and $AA'VO(PO_4)_2$, which are previously studied as spin-1/2 $J_1$–$J_2$ square lattice magnets.

The magnetic susceptibility of $KMoOPO_4Cl$ was reported in the previous study, but was not well characterized [27]. We synthesized single crystals of both the K and Rb compounds and performed magnetization and heat capacity measurements, which are well reproduced by the spin-1/2 $J_1$–$J_2$ square lattice model. Moreover, magnetic transitions are observed at around 6 and 8 K in the K and Rb compounds, respectively. Single crystal NMR experiments on the Rb compound and powder neutron diffraction experiments on both of the compounds show that the ground state is the CAF order in each compound, which must be stabilized by the dominant antiferromagnetic $J_2$. Furthermore, a comparison between the two compounds reveals that the magnitude of the magnetic interactions is sensitive to chemical pressure, which suggests a possible fine tuning of the $J_1$–$J_2$ ratio by physical or chemical pressure to approach to the quantum phases at the boundaries.

II.     EXPERIMENTS

Single crystals of $AMoOPO_4Cl$ ($A$ = K, Rb) were grown by the flux method. $ACl$, NaCl, Mo, $MoO_3$, and $P_2O_5$ powders were mixed in the ratio of 90 : 90 : 1 : 5 : 3 in an Ar-filled glove box. The mixture was put into a Pt tube and sealed in an evacuated quartz tube. The tube was heated to 700 °C at 100 °C/hour and then slowly cooled to 600 °C at 2 °C/hour. Yellow (K) and orange (Rb) square plate crystals of the maximum size of 1 × 1 × 0.2 mm were obtained (insets in Fig.3). They are stable in air at room temperature. Polycrystalline samples for powder neutron diffraction experiments were prepared by the solid state reaction from stoichiometric mixtures heated at 600 and 650 °C for 72 hours for the K and Rb compounds, respectively.

The crystal structure was refined by means of single crystal X-ray diffraction (XRD) using the Shelx software [28]. The crystal structures of the two compounds at 293 K are identical to those reported in the previous study; the lattice constants are $a$ = 6.4362(3) Å and $c$ = 7.2705(6) Å in the K compound and $a$ =

6.4586(3) Å and $c$ = 7.4590(4) Å in the Rb compound, which are close to the values reported in the previous study. A low-temperature XRD experiment was also performed at 90 K for the K compound, because there were signs of a structural distortion in magnetic susceptibility and NMR measurements on the powder sample. Magnetization measurements were performed in a SQUID magnetometer (MPMS3, Quantum Design), and heat capacity measurements were performed by a commercial measurement system by the relaxation method (PPMS, Quantum Design).

To determine the magnetic structure, $^{31}$P NMR ($^{31}\gamma$ = 17.235 MHz) measurements were carried out on the Rb compounds. NMR spectra were obtained by summing the Fourier transform of spin-echo signals obtained at equally spaced rf frequencies with a fixed magnetic field. Neutron powder diffraction measurements were carried out on both compounds using the D1B high-intensity diffractometer at Institute Laue-Langevin [29]. Approximately 5 g of powder samples were loaded into 8 mm diameter V cans in an Ar-filled glove box. The incident neutron wavelength was fixed to $\lambda$ = 2.52 Å. A Rietveld refinement of the nuclear and magnetic structures was performed using the Fullprof package [30].

III.  RESULTS AND DISCUSSION

A.  LOW-TEMPERATURE CRYSTAL STRUCTURE

The crystal structure of the Rb compound remains the same upon cooling, whereas that of the K compound changes below 110 K. A structural analysis on the XRD data obtained at 90 K for the K compound reveals the monoclinic structure with the space group $C2/m$ and the lattice constants $a$ = 9.016(10) Å, $b$ = 9.021(10) Å, $c$ = 7.2468(8) Å and $\beta$ = 90.057(2)°; atomic positions are listed in the Table 1. The low-temperature monoclinic unit cell corresponds to a $\sqrt{2} \times \sqrt{2} \times 1$ superlattice of the high-temperature tetragonal unit cell (Fig. 2(b)). However, the monoclinic distortion is quite small as evidenced by the negligible difference between the $a$- and $b$- axis lengths and the tiny deviation from 90° in $\beta$. In addition, the changes in the bond lengths and angles of the Mo–O bonds are also small. From the structural point of view, we expect that this structural transition does not cause large changes in magnetic interactions. However, it is revealed from the magnetic susceptibility measurements shown below that sizable change in magnetic interactions occur at the transition. Since it occurs only in the K compound, the origin must be related to the size effect arising from the different ionic radii of the $A$ ions, that is, the size mismatch effect or the chemical pressure effect. Possibly, the size of the common layer made of $MoO_5Cl$ octahedra and $PO_4$ tetrahedra fits well with the large Rb ions, but not with the small K ions. This may be related to the lattice distortion caused by the size mismatch governed by the tolerance factor in the perovskite oxides.

B.  MAGNETIC INTERACTIONS

Figures 3 show the temperature dependences of the magnetic susceptibility $\chi$ measured in a magnetic field of 1 T applied along respectively the $a$ and $c$ axis of one single crystal for each compound. At first glance, there is little difference between the two compounds: upon cooling the $\chi$ shows a Curie–Weiss increase, a broad peak at $T_p$, indicating a development of short-range magnetic correlations typical for low-dimensional quantum antiferromagnets, and an anomaly at $T_m$: we define $T_m$ as the temperature of the characteristic minimum in $\chi$ at $H // c$. This confirms that they are essentially the same spin systems. Major differences are found in the values of $T_p$ and $T_m$: ($T_p$ / K, $T_m$ / K) = (21, 6.5) and (26, 8.1) for the K and Rb compounds, respectively. The fittings by the Curie-Weiss law including temperature independent term $\chi_0$, $\chi = \chi_0 + C/(T + \Theta)$ where $C$ and $\Theta$ are the Curie constant and the Weiss temperature, are performed above 150 K. The fittings yield positive Weiss temperatures of 26(1) and 35(1) K for the K and Rb compounds, respectively, indicating predominant antiferromagnetic interactions that seems to scale with the magnitude of $T_p$. The $\chi_0$ terms take tiny negative values between $-4\times10^{-5}$ and $-1\times10^{-4}$ cm$^3$ mol-Mo$^{-1}$ and are considered to represent diamagnetic contributions from core electrons and varnish used to fix crystals; the same $\chi_0$ values are used for the further fittings below. Thus, the overall magnetic interactions are smaller in the K compound than in the Rb compound. The effective magnetic moments estimated from the Curie constants for $H // a$ and $c$ are 1.73(1) and 1.71(1) $\mu_B$ / Mo in the K compound, while 1.67(1) and 1.67(1) $\mu_B$ in the Rb-compound, respectively. These values are quite close to the spin-only value of 1.73 $\mu_B$ for spin 1/2. Therefore, the orbital angular momentum is almost quenched, because the crystal field of the MoO$_5$Cl octahedron has removed the degeneracy of the $t_{2g}$ orbitals completely. The $d_{xy}$ orbital of the Mo$^{5+}$ ion is considered to carry spin-1/2, as that of the V$^{4+}$ ion in the vanadates.

In order to estimate the magnitude of magnetic interactions more precisely, fittings of the magnetic susceptibility data to the high-temperature series expansion for the spin-1/2 Heisenberg square lattice magnet [31] are performed. In the case of the Rb compound, the experimental data are almost completely reproduced down to around 25 K, as shown by the green dashed line in Fig. 3(b): the fitting above 25 K yields an antiferromagnetic interaction of 29(1) K. The $J_1$–$J_2$ square lattice model [15] is also employed but do not improve the fitting. These results suggest that either of $J_1$ or $J_2$ is dominant: $J_1 \sim 29$ K $>>$ |$J_2$| or $J_2 \sim 29$ K $>>$ |$J_1$|. Note that the former case corresponds to the conventional square lattice antiferromagnet, while, in the latter case, two almost independent square lattices are interpenetrating with each other.

In the K compound [Fig. 3(a)], a fitting to the square lattice model at above 120 K for the high-temperature phase gives an antiferromagnetic interaction of 24(1) K, which is slightly smaller than that of the Rb compound. Fitting to the $J_1$–$J_2$ square lattice model did not give reliable results in this temperature range: the result strongly depends on the initial parameter. The apparent enhancement of the magnetic susceptibility from the fitting curve in the low temperature region clearly indicates that the lattice distortion caused by the structural phase transition has modified the magnetic interactions. Probably, the low-temperature structure has smaller antiferromagnetic interactions and/or larger ferromagnetic interactions. The low temperature susceptibility data at 20-100 K can be fitted better to the spin-1/2 $J_1$–$J_2$ square lattice model than to the normal square lattice model, which yields $J_1 = -2.0(1)$ K and $J_2 = 19(1)$ K [the orange dashed line in Fig. 2(a)]. The reverse case of $J_1 = 19$ K and $J_2 = -2.0$ K gives a similarly good fit, which has been excluded by the NMR and the neutron diffraction experiments mentioned below.

C. MAGNETIC TRANSITIONS

The magnetic susceptibilities show anomalies at around $T_m$ in the K and Rb compounds, respectively, below which a distinct anisotropy appears at low fields. Corresponding to these, small kinks are observed in heat capacity at similar temperatures, as shown in Fig. 4, indicating long-range magnetic orders at $T_m$. Concerning the anisotropy, the susceptibility decreases in $H // a$, while increase in $H // c$ at 1 T. At 2 T, the susceptibility also shows an upturn in $H // a$, but the difference between the $H // a$ and $c$ measurements still remains with a reduced anisotropy. The difference completely vanishes at 5 T, indicating an isotropic magnetic order. The $T_m$ increases gradually with increasing $H$ in both compounds.

The heat capacity data of the two compounds are also similar to each other. At zero field it shows a tiny kink at 6.3 (7.8) K and a broad shoulder at around 12 (15) K for the K (Rb) compound. With increasing magnetic field, the kink shifts to higher temperatures up to 7.4 (9.0) K at 9 T and tends to become sharp. We define these anomaly temperatures as the magnetic ordering temperatures $T_N$ and plot them in the $H$-$T$ diagram of Fig. 5. The small anomalies, particularly at $H = 0$, indicate that the magnetic entropy released at $T_N$ is small, which is attributed to the good two-dimensionality of the spin systems: most entropy is released by short-range correlations at higher temperatures, which is actually observed as a broad peak in the heat capacity.

The observed temperature dependences in magnetic susceptibility are quite similar to those calculated for the spin-1/2 Heisenberg square lattice antiferromagnet with small easy-plane anisotropy [10]. For example, in the presence of easy-plane anisotropy of 0.02$J$, the calculated out-of-plane susceptibility shows a minimum at $T \sim 0.3J$, which is considered to represent a crossover from isotropic to XY spin behavior, preceding Berezinskii–Kosterlitz–Thouless transition at $T \sim 0.23J$. The Rb compound with $J_2 \sim$ 29 K, shows a minimum at $T_m \sim 0.28J_2$ in $H // c$. This coincidence suggests the presence of easy-plane anisotropy of the order of 0.02$J_2$. Note that $T_N$ determined from heat capacity is slightly lower than $T_m$: $T_N \sim 0.27J_2$. Increase of the $T_N$ in $H // c$ should be attributed to the enhancement of effective easy plane anisotropy by the applied magnetic fields [32]. Origin of the growth of the peak in heat capacity is not clear, but may be related to a field-induced spin-flop transition or change in the type of magnetic transition.

D. $^{31}$P NMR RESULTS

Microscopic information about the magnetic order of the Rb compound has been examined by $^{31}$P NMR experiments. Figures 6 (a) and (b) show the temperature dependences of the NMR spectra measured at 5 T in $H // c$ and $ab$, respectively. In each case, a sharp single peak is observed above 9 K, which is consistent with the fact that the P atoms occupy a unique crystallographic site. The magnetic shift in the paramagnetic phase should follow a uniaxial angular dependence, because the P site has the $S_4$ symmetry around the $c$ axis [Fig. 6(c)]. The hyperfine coupling constants are determined to be $A_{//} = -0.43$ T/$\mu_B$ and $A_{\perp} = -0.09$ T/$\mu_B$ from the data of the magnetic shift and the susceptibility, which show a linear relation in the paramagnetic phase. The single paramagnetic peak in $H // c$ splits into two peaks below 8 K, as shown in Fig. 6(a), which indicates the appearance of an antiferromagnetic internal field. The temperature dependence of the internal field $H_{int}$ determined from the peak splitting is shown in Fig. 6(d). By fitting the data to the function $H_{int} = H_0(1 - T/T_N)^\beta$, we obtain the transition temperature $T_N = 8.12(1)$ K at 5 T and the critical exponent $\beta = 0.24(1)$. The field dependence of $T_N$ from NMR is plotted in Fig. 5, which coincides with that from heat capacity. Note that, in addition to the split peaks, another peak that could be attributed to the paramagnetic state is observed in the narrow temperature range of 8.0–8.1 K. This coexistence may be attributed to inhomogeneity due to certain imperfection of the crystal or a weak first-order nature of the transition.

In sharp contrast, the NMR spectra in $H // ab$ do not show such a splitting as observed in $H // c$ but a slight broadening below 8 K [Fig. 6(b)]. This anisotropic behavior gives a strong constraint on the magnetic structure. According to the calculations of the hyperfine field on the P site generated by magnetic moments on the four surrounding Mo sites given in Appendix, the NMR spectrum in $H // c$ should have a single peak in the NAF state, whereas it shows a double peak in the CAF state except for the case that the magnetic moments point to the [110] direction. Therefore, it is concluded that the magnetic structure is a CAF state with its magnetic moment not parallel to the [110] direction. On the other hand, the calculations also indicate that the internal field perpendicular to the $c$ axis is proportional to the ordered moment parallel to the $c$ axis in the CAF state. Thus, no splitting is expected in the NMR spectrum in $H // ab$, when the magnetic moments lie within the $ab$ plane. The slight broadening observed in $H // ab$ must indicate that the ordered moments lie nearly in the $ab$ plane.

E.  POWDER NEUTRON DIFFRACTION

At base temperature of 1.5 K, we observe a number of magnetic Bragg peaks which can be indexed by a single propagation wave vector of $k = (0.5, 0.5, 0.5)$ in each compound, as shown in Fig. 7. They appear below ~ 7 and ~ 8 K in the K and Rb compound, respectively, as shown in the insets. In order to analyze the magnetic structures from neutron diffraction we have performed the decomposition of the magnetic representation into irreducible representation of the paramagnetic space group, that is, $P4/nmm$ in the present case, using the Basireps program in the Fullprof package [30]. We obtain three two-dimensional irreducible representations, $\Gamma_k = \Gamma_2 + \Gamma_3 + \Gamma_4$. The Fourier coefficients $S$ for each irreducible representation are shown in Table II for the two symmetry-related Mo sites. As the propagation wave vector $k$ is equivalent to $-k$, the magnetic moments are found as $m(L_x, L_y, L_z) = S_n(-1)^{(L_x+L_y+L_z)}$, where $L_{x,y,z}$ are real space translation vectors and n labels the zeroth unit cell Mo ion. To approximate the magnetic form factor of the Mo$^{5+}$ ions, which is not available in literature, we take an average of the magnetic forms factors for $Cr^{4+}$ and $W^{5+}$ ions [33-34].

Taking into account the magnetic symmetry and form factor, we perform Rietveld refinements of the magnetic structure with the moment size and directions as free parameters. The refined patterns for the K(Rb) compound have a goodness-of-fit $\chi^2$ = 13.5 (10.9) and magnetic $R$-factor of 48.1 (22.3). As shown in Fig. 7, the main source of the relatively poor fit quality is due to imperfect subtraction of the nuclear contribution as a result of slight lattice contraction on cooling from 30 to 1.5 K. Due to powder averaging we cannot rule out a non-collinear magnetic structure based on our measurements, and cannot determine the moment direction in the *ab* plane. Nevertheless, symmetry analysis allows us to greatly constrain the phases between ions to that of the $\Gamma_4$ irreducible representation which best describes the powder diffraction patterns of both the compounds. An example of a magnetic structure that is consistent with the data is depicted in Fig. 8. It consists of antiferromagnetic chains along [1, 0, 0] that are ferromagnetically stacked along the [1, 1, 0] direction, which is the CAF structure and perfectly consistent with the NMR results. An additional information is on the stacking of sheets: the spins are arranged antiferromagnetically along the *c* axis. The ordered magnetic moment on the Mo ion is within fitting uncertainty the same in both compounds, 0.54(6) $\mu_B$ in $KMoOPO_4Cl$ and 0.53(1) $\mu_B$ in $RbMoOPO_4Cl$. The large reduction of the ordered moments is likely to originate from quantum fluctuation.

F.  COMPARISON BETWEEN THE $J_1$–$J_2$ SQUARE LATTICE COMPOUNDS

The present NMR and neutron diffraction experiments reveal CAF orders for both the compounds. To reconcile this, $J_2$ must be larger than $J_1$. Thus, the magnetic interactions of the K and Rb compounds are determined as ($J_1$ / K, $J_2$ / K) = (−2, 19) and (~ 0, 29) for the K and Rb compounds, respectively; the $J_2$ is slightly larger and the $T_N$ is higher in the Rb compound than in the K compound. Consequently, the two compounds provide similar $J_1$–$J_2$ square lattice antiferromagnets with relatively large antiferromagnetic $J_2$ and much smaller $J_1$, so that they exist near the top of the $J_1$–$J_2$ phase diagram of Fig. 1.

A small but significant difference between the two compounds comes from the $J_1$ values. Because of the negligible $J_1$, the Rb compound provides a unique antiferromagnet made of weakly coupled interpenetrating square lattices. For $J_1 \sim 0$, one naively expects two independent square lattices, because the nearest-neighbor couplings are geometrically cancelled. However, they prefer the CAF long-range magnetic order by the order-by-disorder mechanism [1]. If the $J_1$ is neglected, this compound may be regarded as an ideal model compound of the spin-1/2 square lattice antiferromagnet. As evidenced from the anisotropy in magnetic susceptibility, it is a nearly Heisenberg magnet with small easy-plane anisotropy. It may offer an ideal playground to study quantum magnetism such as a quantum effect on the magnetic excitations [11].

On the other hand, the K compound at low temperatures below structural transition has $J_1 \sim -2$ K. The $J$ = 24 K estimated above the structural transition is considered to come from mostly $J_2$ as in the Rb compound without structural transition. Then, the lattice distortion has enhanced ferromagnetic $J_1$ and reduced antiferromagnetic $J_2$. Although the lattice distortion associated with the structural transition is very small, the superexchange path via the $PO_4$ unit must be highly sensitive to a change in the local structure, which is caused by the chemical pressure effect. The tendency that the chemical pressure acts to decrease the $J_2/J_1$ ratio coincides with the calculation performed for the crystal structures of the $3d$ $J_1$–$J_2$ square lattice antiferromagnet $Li_2VOSiO_4$ under high pressure [35]. Interestingly, in the structurally related compound $MoOPO_4$, a Néel-type magnetic structure, where the moments arranged parallel to the $c$-axis in the plane and stacked ferromagnetically, is observed [36]. Note that, owing to the "de-intercalation" of $A$ and Cl ions from $A$MoOPO$_4$Cl, the $MoO_6$ octahedra and $PO_4$ tetrahedra tilt mutually in the $ab$-plane in $MoOPO_4$, which might result in a dominant antiferromagnetic $J_1$ to the NAF order. This comparison suggests that magnetic interactions in the $MoOPO_4$ layer are so sensitive to the crystal structure as to change the ground state magnetic structure.

The CAF order has been observed in the $3d$ compounds $Li_2VOMO_4$ ($M$ = Si, Ge) and $AA'$VO(PO$_4$)$_2$ ($AA'$ = Pb$_2$, SrZn, BaZn, and BaCd). Note that these vanadates have $|J_1|$ and $|J_2|$ smaller than 10 K. The larger antiferromagnetic interaction in the molybdates must be explained by the larger overlapping between more spatially more extended $4d$ orbitals and O $2p$ orbitals, which increases the electron transfer $t$ associated with the superexchange process. Moreover, the smaller electron correlation $U$ of the Mo ion may also enhance the magnetic coupling; $J \sim t^2/U$.

IV. SUMMARY

We have studied the magnetism of $A$MoOPO$_4$Cl ($A$ = K, Rb) with the $4d^1$ electronic configuration and showed that the compounds provide us with good candidates for the spin-1/2 $J_1$–$J_2$ square lattice magnet. The magnetic interactions are estimated as $|J_1| \sim 0$ and $J_2 \sim 29$ K in the Rb compound and $J_1 \sim -2$ K and $J_2 \sim 19$ K in the K compound. The magnetic structure of each compound is determined by the NMR and neutron diffraction experiments, which is a CAF order as expected for the corresponding parameter regime from the theoretical prediction for the $J_1$–$J_2$ square lattice magnet. In addition, magnetic moments with reduced magnitude of ~0.5 $\mu_B$ lie within the plane. On the other hand, we have found that the magnetic interactions are sensitive to the local crystal structures. Applying further chemical or physical pressure would increase $J_1$ and decrease $J_2$ to drive the system closer to the highly frustrated parameter region. NMR experiments under pressure are now ongoing.

ACKNOWLEDGMENTS

HI was supported by research fellowship of Japan Society for the Promotion of Science. HMR thanks ISSP for hospitality. We wish to thank C. Ritter for his insight in estimating the Mo$^{5+}$ magnetic form factor and M. Jeong and O. Janson for helpful discussions. This work was partially funded by the European Research Council grant CONQUEST the Swiss National Science Foundation and its Synergia network MPBH.

[1] G. Misguich and C. Lhuillier, in *Frustrated Spin Systems*, edited by H. T. Diep (World-Scientific, Singapore, 2005), p. 229.

[2] H.-C. Jiang, H. Yao, and L. Balents Phys. Rev. B 86, 024424 (2012).

[3] S.-S. Gong, W. Zhu, D. N. Sheng, O. I. Motrunich, and M. P. A. Fisher, Phys. Rev. Lett. 113, 027201 (2014).

[4] N. Shannon, T. Momoi, and P. Sindzingre, Phys. Rev. Lett. 96, 027213 (2006).

[5] Q. Si, and E. Abrahams. Phys. Rev. Lett. 101, 076401 (2008).

[6] T. Yildirim, Phys. Rev. Lett. 101, 057010 (2008).

[7] D. Vaknin, S. K. Sinha, D. E. Moncton, D. C. Johnston, J. M. Newsam, C. R. Safinya, and H. E. King, Jr. Phys. Rev. Lett. 58, 2802 (1987).

[8] R. Coldea, S. M. Hayden, G. Aeppli, T. G. Perring, C. D. Frost, T. E. Mason, S.-W. Cheong, and Z. Fisk, Phys. Rev. Lett. 86, 5377 (2001).

[9] M. Greven, R. J. Birgeneau, Y. Endoh, M. A. Kastner, M. Matsuda, G. Shirane, Zeitschrift für Physik B Condensed Matter 96.465 (1995).

[10] A. Cuccoli, T. Roscilde, R Vaia, and P. Verrucchi, Phys. Rev. Lett. 90, 167205 (2003).

[11] B. D. Piazza, M. Mourigal, N. B. Christensen, G. J. Nilsen, P. Tregenna-Piggott, T. G. Perring, M. Enderle, D. F. McMorrow, D. A. Ivanov, and H. M. Rønnow, Nat. Phys. 11, 62 (2015).

[12] N. Tsyrulin, F. Xiao, A. Schneidewind, P. Link, H. M. Rønnow, J. Gavilano, C. P. Landee, M. M. Turnbull, and M. Kenzelmann, Phys. Rev. B 81, 134409 (2010).

[13] J. L. Manson, M. M. Conner, J. A. Schlueter, T. Lancaster, S. J. Blundell, M. L. Brooks, F. L. Pratt, T. Papageorgiou, A. D. Bianchi, J. Wosnitzae, and M.-H. Whangboo, Chem. Commun. 4894 (2006).

[14] R. Melzi, S. Aldrovandi, F. Tedoldi, P. Carretta, P. Millet, and F. Mila, Phys. Rev. B 64, 024409 (2001).

[15] H. Rosner, R. R. P. Singh, W. H. Zheng, J. Oitmaa, S.-L. Drechsler, and W. E. Pickett, Phys. Rev. Lett. 88, 186405 (2002).

[16] H. Rosner, R. R. P. Singh, W. H. Zheng, J. Oitmaa, and W. E. Pickett, Phys. Rev. B 67, 014416 (2003).

[17] A. Bombardi, L. C. Chapon, I. Margiolaki, C. Mazzoli, S. Gonthier, F. Duc, and P. G. Radaelli Phys. Rev. B 71, 220406(R) (2005).

[18] A. A. Tsirlin, A. A. Belik, R. V. Shpanchenko, E. V. Antipov, E. Takayama-Muromachi, and H. Rosner, Phys. Rev. B 77, 092402 (2008).

[19] R. Nath, A. A. Tsirlin, H. Rosner, and C. Geibel, Phys. Rev. B 78, 064422 (2008).

[20] E. E. Kaul, H. Rosner, N. Shannon, R. V. Shpanchenko, C. Geibel, J. Mag. Mag. Mat., 272, 922 (2004).

[21] A. A. Tsirlin, and H. Rosner, Phys. Rev. B 79, 214417 (2009).

[22] A. A. Tsirlin, R. Nath, A. M. Abakumov, R. V. Shpanchenko, C. Geibel, and H. Rosner, Phys. Rev. B 81, 174424 (2010).

[23] I. De, V. P. Desai, and A. S. Chakravarty, Phys. Rev. B 8, 3769 (1973).

[24] E. E. Rodriguez, F. Poineau, A. Llobet, B. J. Kennedy, M. Avdeev, G. J. Thorogood, M. L. Carter, R. Seshadri, D. J. Singh, and A. K. Cheetham, Phys. Rev. Lett. 106, 067201 (2011).

[25] G. Chen, R. Pereira, and L. Balents, Phys. Rev. B 82, 174440 (2010).

[26] J. A. Sears, M. Songvilay, K. W. Plumb, J. P. Clancy, Y. Qiu, Y. Zhao, D. Parshall, and Y.-J. Kim Phys. Rev. B 91, 144420 (2015).

[27] M. M. Borel, A. Leclaire, J. Chardon, J. Provost, B. Raveau J. Solid State Chem. 137 214 (1998).

[28] G. M. Sheldrick, Acta Cryst. C71, 3 (2015).

[29] P. Babkevich, Z. Hiroi, N. Qureshi, and H.M. Ronnow, Magnetic order in potential J1–J2 square lattice Heisenberg system. Institut Laue-Langevin (ILL) (2015) doi:10.5291/ILL-DATA.5-31-2401

[30] J. Rodriguez-Carvajal, Physica B 192, 55 (1993).

[31] G. S. Roshbrooke, and P. J. Wood, Molecular Physics 1 257 (1958).

[32] A. Cuccoli, T. Roscilde, R. Vaia, and P. Verrucchi, Phys. Rev. B 68, 060402(R) (2003).

[33] We employ the standard analytical approximation of the magnetic form factor, $f = A \exp(-as^2) + B \exp(-bs^2) + C \exp(-cs^2) + D$, where $s = \sin\theta / \lambda$. From fitting the averaged $Cr^{4+}$ and $W^{5+}$ magnetic form factor [34], we find $A = -0.508$, $a = -0.025$, $B = 0.356$, $b = 21.573$, $C = 0.626$, $c = 11.707$, and $D = 0.524$.

[34] K. Kobayashi, T. Nagao, and M. Ito, Acta Crystalographica A**67**, 473 (2011).

[35] E. Pavarini, S. C. Tarantino, T. B. Ballaran, M. Zema, P. Ghigna, and P. Carretta, Phys. Rev. B. **77**, 014425 (2008).

[36] L. Yang, M. Jeong, P. Babkevich, V. M. Katukuri, B. Nafradi, A. Magrez, H. Berger, J. Schefer, O. Zaharko, M. Kriener, I. Zivkovic, O. V. Yazyev, L. Forro, and H. M. Rønnow (unpublished).

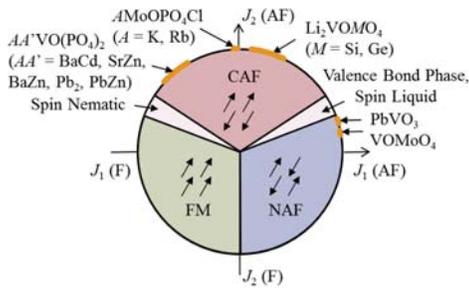

FIG 1. Ground state phase diagram of the spin-1/2 $J_1$–$J_2$ square lattice magnet and the model compounds placed on it. NAF, FM, and CAF refer to the Néel antiferromagnetic, the ferromagnetic, and the columnar antiferromagnetic states, respectively. Quantum ground states without long-range magnetic orders are predicted near the boundaries between NAF and CAF or between FM and CAF.

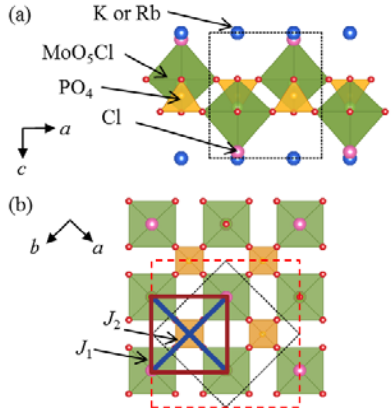

FIG 2. Crystal structures of $A$MoOPO$_4$Cl ($A$ = K, Rb) viewed along the $b$ axis (a) and the $c$ axis (b). MoO$_5$Cl octahedra (green), PO$_4$ tetrahedra (yellow), and spheres of K or Rb ions (blue) are depicted. The red and purple spheres represent oxide and chloride ions, respectively. The unit cell of the $P4/nmm$ structure and that of the low-temperature $C2/m$ structure of the K compound are shown by the black and red dashed lines, respectively. The NN interaction $J_1$ and the NNN interaction $J_2$ are depicted by the thick red and blue lines in (b), respectively.

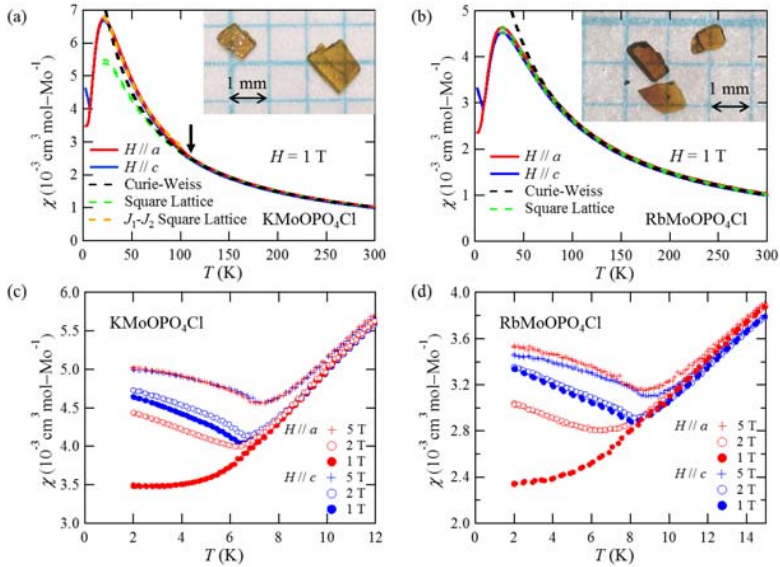

FIG 3. Temperature dependences of magnetic susceptibility for KMoOPO$_4$Cl (a, c) and RbMoOPO$_4$Cl (b, d) measured. For each measurement, one single crystal such as shown in the inset photograph was used. The red and blue lines show the data in the magnetic field $H // a$ and $c$, respectively. The black dashed lines in (a) and (b) show Curie–Weiss fits, and the green and orange dashed lines show fittings to the high-temperature series expansions of the spin 1/2 square-lattice magnet and the $J_1$–$J_2$ square lattice magnet, respectively. The black arrow in (a) for the K compound indicates the temperature of the structural phase transition, which is absent in the Rb compound. The magnetic susceptibility data measured in different magnetic fields of 1, 2, and 5 T at low temperatures are shown in (c) and (d).

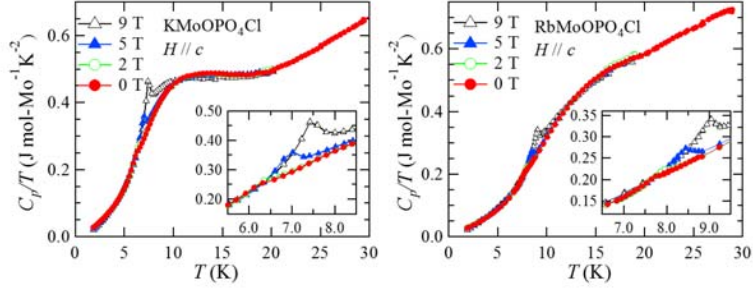

FIG 4. Temperature dependences of heat capacity for (a) KMoOPO$_4$Cl and (b) RbMoOPO$_4$Cl in the magnetic fields of 0 T (red filled circle), 2 T (green open circle), 5 T (blue filled triangle), and 9 T (black open triangle) along the $c$ axis. Enlarged plots around the anomalies are shown in the inset.

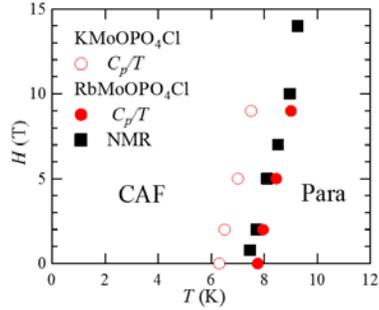

FIG 5. Temperature-magnetic field phase diagram of $A$MoOPO$_4$Cl ($A$ = K, Rb) obtained by the heat capacity and NMR measurements in the magnetic field $H \mathbin{/\mkern-2mu/} c$. Para and CAF represent the paramagnetic and columnar antiferromagntic ordered states, respectively.

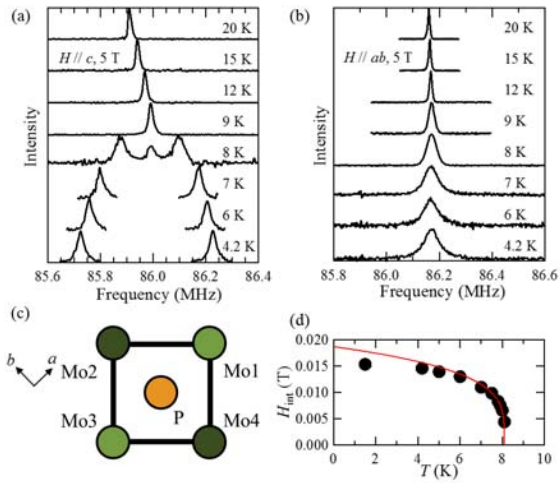

FIG 6. Temperature dependences of the $^{31}$P NMR spectra for a single crystal of RbMoOPO$_4$Cl at 5 T in (a) $H \text{ // } c$ and (b) $H \text{ // } ab$. (c) Coordination around a P atom viewed along the $c$ axis in RbMoOPO$_4$Cl. Four Mo atoms are arranged in the $S_4$ symmetry: Mo1 and Mo3 atoms are located at a height below the P atom, and Mo2 and Mo4 atoms above. (d) Temperature dependence of the internal field $H_{int}$ determined from the peak splitting at 5 T in $H \text{ // } c$. The red line is a fit to the form $H_{int} = H_0(1 - T/T_N)^\beta$.

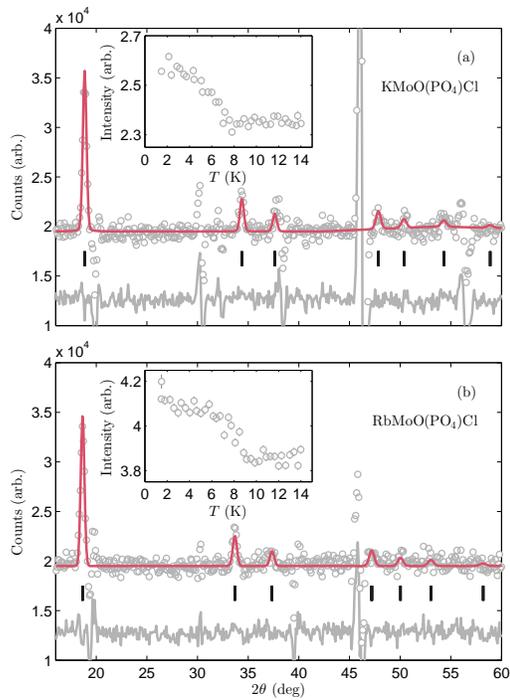

FIG. 7. Neutron powder diffraction patterns collected at 1.4 K for (a) KMoOPO$_4$Cl and (b) RbMoOPO$_4$Cl. For each data, a paramagnetic background from 30 K measurements has been subtracted. Plotted in red is the calculated pattern according to the magnetic structure model described in the text. Black vertical lines denote the magnetic Bragg peak index. The difference between simulations and measured intensities is shown at the bottom of each panel. The temperature dependence of the integrated intensity of the (0.5, 0.5, 0.5) reflection is shown in the inset for each compound.

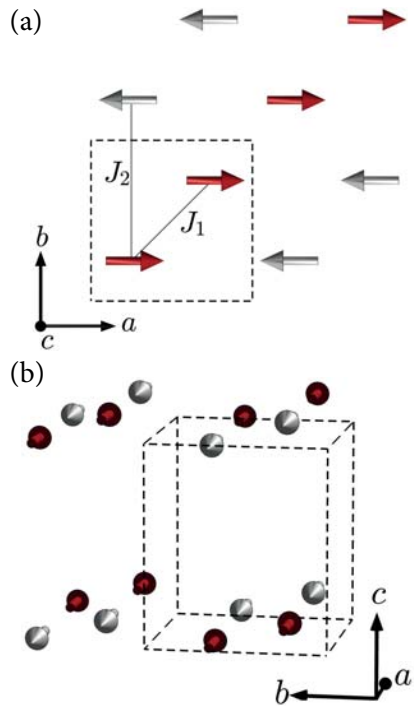

FIG 8. Possible magnetic structure of $A$MoOPO$_4$Cl ($A$ = K, Rb) determined by the powder neutron experiments.

TABLE 1. Atomic parameters obtained by the single crystal XRD experiments of KMoOPO$_4$Cl at 90 K. The lattice constants are $a$ = 9.016(10) Å, $b$ = 9.021(10) Å, $c$ = 7.2468(8) Å and $\beta$ = 90.057(2)° in the monoclinic space group $C2/m$.

| Atom | Site | $x$ | $y$ | $z$ | $100U_{eq}$ (Å$^2$) |
|---|---|---|---|---|---|
| K | 4g | 0.5 | 0.2501(2) | 0 | 1.43(3) |
| Mo | 4i | 0.7500(1) | 0 | 0.4054(1) | 0.73(2) |
| P | 4h | 0.5 | 0.2500(2) | 0.5 | 0.71(3) |
| O1 | 4i | 0.7500(6) | 0 | 0.6321(7) | 1.7(1) |
| O2 | 8j | 0.5926(4) | 0.3442(4) | 0.6303(4) | 1.3(1) |
| O3 | 8j | 0.5924(4) | 0.1558(4) | 0.3697(4) | 1.3(1) |
| Cl | 4i | 0.7499(2) | 0 | 0.0488(2) | 1.06(3) |

TABLE 2. Fourier coefficients $S$ of irreducible representation $\Gamma\nu$ separated into real and imaginary components and resolved along the crystallographic axes. Mo ions are situated at 1. (0.25, 0.25, 0.09) and 2. (−0.25, 0.75, −0.09). The coefficients $u$ and $v$ are free parameters which must be determined experimentally.

| $\nu$ | | Mo$_1$ | Mo$_2$ |
|---|---|---|---|
| 2 | Re | (0, 0, $u$) | (0, 0, −$v$) |
|   | Im | (0, 0, −$v$) | (0, 0, $u$) |
| 3 | Re | ($u$, −$v$, 0) | (−$v$, $u$, 0) |
|   | Im | (−$v$, $u$, 0) | ($u$, −$v$, 0) |
| 4 | Re | ($u$, $v$, 0) | (−$v$, −$u$, 0) |
|   | Im | (−$v$, −$u$, 0) | ($u$, $v$, 0) |

APPENDIX. CALCULATION OF THE HYPERFINE FIELD

We discuss the magnetic structure of RbMoOPO$_4$Cl based on the symmetry characteristic of the hyperfine coupling tensor. Since the hyperfine interaction in insulators are short-ranged, we can assume that the hyperfine field at the P sites is the sum of the contributions from the four nearest-neighbor Mo sites,

$$\mathbf{H}_{\text{int}} = \sum_{i=1}^{4} \mathbf{A}_i \cdot \mathbf{m}_i, \quad (\text{A.1})$$

where $\mathbf{A}_i$ is the hyperfine coupling tensor between the P nucleus and the magnetic moment $\mathbf{m}_i$ of the $i$-th Mo [see Fig. 5(c)]. Since the Mo1 and P sites are on the mirror plane perpendicular to the $b$ axis, it is one of the principal axes of $\mathbf{A}_1$. Therefore, $\mathbf{A}_1$ can be expressed as

$$\mathbf{A}_1 = \begin{pmatrix} A_{aa} & 0 & A_{ac} \\ 0 & A_{bb} & 0 \\ A_{ac} & 0 & A_{cc} \end{pmatrix}. \quad (\text{A.2})$$

Then, $\mathbf{A}_2$, $\mathbf{A}_3$, and $\mathbf{A}_4$ are obtained by applying $S_4$ symmetry operation,

$$\mathbf{A}_2 = \begin{pmatrix} A_{bb} & 0 & 0 \\ 0 & A_{aa} & -A_{ac} \\ 0 & -A_{ac} & A_{cc} \end{pmatrix}, \quad \mathbf{A}_3 = \begin{pmatrix} A_{aa} & 0 & -A_{ac} \\ 0 & A_{bb} & 0 \\ -A_{ac} & 0 & A_{cc} \end{pmatrix}, \quad \mathbf{A}_4 = \begin{pmatrix} A_{bb} & 0 & 0 \\ 0 & A_{aa} & A_{ac} \\ 0 & A_{ac} & A_{cc} \end{pmatrix}. \quad (\text{A.3})$$

First, we consider the NAF order, where the ordered moments are described as

$$\mathbf{m}_1 = -\mathbf{m}_2 = \mathbf{m}_3 = -\mathbf{m}_4 = \boldsymbol{\sigma}^{\text{I}} = \begin{pmatrix} \sigma_a^{\text{I}} \\ \sigma_b^{\text{I}} \\ \sigma_c^{\text{I}} \end{pmatrix}. \quad (\text{A.4})$$

By incorporating Eqs. (A.2), (A.3), and (A.4) into Eq. (A1), we obtain the internal field

$$\mathbf{H}_{\text{int}} = 2(A_{aa} - A_{bb}) \begin{pmatrix} \sigma_a^{\text{I}} \\ -\sigma_b^{\text{I}} \\ 0 \end{pmatrix}. \quad (A.5)$$

Thus, the internal field along the $c$ axis is zero in the NAF order. Furthermore, the $ab$ components of $\mathbf{H}_{\text{int}}$ are negligible in $H$ // $c$, when the applied magnetic field is sufficiently strong, $H \gg |\mathbf{H}_{\text{int}}|$, which is the case in our experiments. Therefore, in the NAF, the NMR spectrum in $H$ // $c$ must show a single peak, which is inconsistent with our NMR results. In contrast, the ordered moments in the CAF order are described as

$$\mathbf{m}_1 = \mathbf{m}_2 = -\mathbf{m}_3 = -\mathbf{m}_4 = \boldsymbol{\sigma}^{\text{II}} = \begin{pmatrix} \sigma_a^{\text{II}} \\ \sigma_b^{\text{II}} \\ \sigma_c^{\text{II}} \end{pmatrix}. \quad (A.6)$$

By incorporating Eqs. (A.2), (A.3), and (A.6) into Eq. (A1), we obtain

$$\mathbf{H}_{\text{int}} = 2A_{ac} \begin{pmatrix} \sigma_c^{\text{II}} \\ -\sigma_c^{\text{II}} \\ \sigma_a^{\text{II}} - \sigma_b^{\text{II}} \end{pmatrix}. \quad (A.7)$$

In this case, the internal field has a $c$ component except for a special case where the magnetic moment is pointing to the diagonal direction between the $a$ and $b$ axes. Since there is another P site which feels internal field with opposite sign in the CAF structure, two peaks are expected in the NMR spectrum for $H$ // $c$, which is consistent with our experiment. Moreover, the absence of splitting in $H$ // $ab$ indicates that $\sigma_c^{\text{II}}$ is almost zero, i.e., the magnetic moment lies in the $ab$ plane.